\documentclass[journal]{IEEEtran}
\ifCLASSINFOpdf
\else
\fi
\usepackage{makecell}

\usepackage{bbding}
\usepackage{graphicx}
\usepackage{amsmath}
\usepackage{amssymb}
\usepackage{float}
\usepackage{stfloats}
\usepackage{algorithm}
\usepackage{algpseudocodex}
\usepackage{cite} 
\usepackage{bm}
\usepackage{soul}
\usepackage{upgreek}
\usepackage{textgreek}
\usepackage{multirow}
\usepackage{multicol}
\usepackage{siunitx}
\usepackage{hyperref}

\hypersetup{
    pdftitle={Neuromorphic Event-Driven Semantic Communication in Microgrids},    % title
    pdfauthor={},     % author
    pdfsubject={},   % subject of the document
    pdfcreator={},   % creator of the document
    pdfproducer={},  % producer of the document
    pdfkeywords={}, % list of keywords
}

\begin{document}
\bstctlcite{bstctl:nodash}
\title{Neuromorphic Event-Driven Semantic Communication in Microgrids}

\author{Xiaoguang Diao,~\IEEEmembership{Student Member, IEEE},
        Yubo Song,~\IEEEmembership{Member, IEEE}, 
        Subham Sahoo,~\IEEEmembership{Senior Member, IEEE}  and 
        Yuan Li, \IEEEmembership{Student Member, IEEE}
\thanks{This work is supported by the Nordic Energy Research programme via Next-uGrid project n. 117766. The authors would like to acknowledge Prof. Osvaldo Simeone for his advice and inputs in the organization of this paper.

{Xiaoguang Diao is with Hubei Key Laboratory for High-efficiency Utilization of Solar Energy and Operation Control of Energy Storage System, Hubei University of Technology, Wuhan 430068, China and the Department of Energy, Aalborg University, Denmark. (e-mail: 2018202070081@whu.edu.cn)}

Yubo Song, Subham Sahoo and Yuan Li are with the Department of Energy, Aalborg University, Denmark. (e-mail: \{\texttt{yuboso, sssa, yuanli}\}@energy.aau.dk)\\
\textit{Corresponding author: Subham Sahoo}}}
\maketitle

% Outline of Abstract 
%
\begin{abstract}
Synergies between advanced communications, computing and artificial intelligence are unraveling new directions of coordinated operation and resiliency in microgrids. On one hand, coordination among sources is facilitated by distributed, privacy-minded processing at multiple locations, whereas on the other hand, it also creates exogenous data arrival paths for adversaries that can lead to cyber-physical attacks amongst other reliability issues in the communication layer. This long-standing problem necessitates new intrinsic ways of exchanging information between converters through power lines to optimize the system's control performance. Going beyond the existing power and data co-transfer technologies that are limited by efficiency and scalability concerns, this paper proposes neuromorphic learning to implant communicative features using spiking neural networks (SNNs) at each node, which is trained collaboratively in an online manner simply using the power exchanges between the nodes. As opposed to the conventional neuromorphic sensors that operate with spiking signals, we employ an event-driven selective process to collect sparse data for training of SNNs. Finally, its multi-fold effectiveness and reliable performance is validated under simulation conditions with different microgrid topologies and components to establish a new direction in the sense-actuate-compute cycle for power electronic dominated grids and microgrids.
%Reliability and stability are important to the cooperative control of microgrids (MGs), but they can be compromised by communication issues, such as time delays, cyberattacks, and line outages. The existing methods such as Talkative Power Communication (TPC) can't solve such problems since signal modulation and channels are always necessities. To solve this issue, this paper introduces a Neuromorphic Semantic Communication (NSC) methodology for microgrids that utilizes a Spiking Neural Network (SNN) deployed at each converter to gather local samplings. By decoding variations of voltage and current, the target information from other converters can be obtained free from physical communication channels. Besides, NSC is also energy-efficient where the event-driven binary spikes ``0" and ``1" are remarkably sparse in the computing process, compared with the always-on real-valued computing in Artificial Neural Networks (ANNs). In the end, case studies in DC microgrids are presented to demonstrate the feasibility and the advantages of NSC-based control. 
\end{abstract}

\begin{IEEEkeywords}
Neuromorphic computing, microgrids, hierarchical control, spiking neural network, semantic communication.
\end{IEEEkeywords}

% \fi
\IEEEpeerreviewmaketitle

%\vspace{-0.3cm}
\section{Introduction}
\IEEEPARstart
{M}{icrogrids} (MGs) have demonstrated their effectiveness in integrating distributed energy sources and energy storage, such as fuel cells, distributed wind and solar generations, and microturbines with the operation both in grid-connected and islanded mode \cite{sahoo2017distributed}.
To achieve proper coordination and maximum utilization from its components, many hierarchical control strategies including primary, secondary, and tertiary controls are utilized to address short and long term objectives, ranging from system dynamics to economic optimization in the system. 
{As reviewed in \cite{guerrero2012advanced}, primary control operates with inner control of distributed generation (DG) units by adding virtual inertia. However, the lack of access to information from other sources makes it deviate from the nominal operation point}  \cite{sahoo2017adaptive}. On the other hand, upon deploying a communication network for information exchange between converters, secondary controllers compensate for the errors introduced by the primary control layer and achieve coordinated objectives, such as current sharing, voltage regulation, and power sharing. {Objectives can also be specially designed as multiple optimization problems in complex topology as defined in \cite{espina2020consensus} and \cite{espina2023consensus}. The operation cost can be minimized by optimizing the dispatch of DGs.} {Meanwhile, the communication structure is also developed from the centralized structure to a more resilient distributed structure. Going beyond the centralized information collection, distributed control philosophy improves the system robustness by only requiring the states of adjacent nodes for sparsity in achieving system-level convergence.} To this end, {hierarchical coordinated control is a unique context of MGs that is highly dependent on communication.} 
% Distributed secondary control framework is significantly more reliable as they only require the states of adjacent nodes for sparsity in achieving system-level convergence.
{Predictive control in the secondary level in MGs is an alternative for mitigating data dropouts and latency by providing latency compensation and modified adjacency matrix in response to electrical or communications disturbances \cite{mfdc}.}
However, communication networks still expose microgrids to {specific challenges, such as random communication delays \cite{dong2019stability}, cyber attacks \cite{sahoo2018stealth}, and cyber link outages \cite{wu2022line}}.  \par
 \textit{Talkative Power Communication} (TPC) has emerged as an innovative solution \cite{angjelichinoski2016multiuser,he2020nature} for co-transfer of power and information, that transmits messages through power lines by encoding and superimposing on the respective bus voltages. 
 Various digital modulation techniques, such as amplitude-shift keying (ASK), frequency-shift keying (FSK), and phase-shift keying (PSK), are utilized for overlaying data onto a reference signal  \cite{zhu2018embedding}. An alternative approach involves modifying the carrier during the modulation process \cite{chen2022simultaneous}. TPC modulations can also be implemented entirely in the digital domain \cite{hoeher2022networking}.
{However in AC systems, these signals cannot traverse through  transformers due to the absence of a zero-sequence path. Similarly, in DC systems, these signals are obstructed by solid-state medium frequency transformers (SSTs) as SST inherently functions as a DC-DC converter which decouples the input and output. The encoded signals can also be attenuated by the input capacitor of solid-state transformers (SSTs). This constraint restricts the application of TPC in microgrids with different voltage levels, which thereby encounters significant technical difficulties when scaling up to high voltage (HV) levels and number of converters \cite{leng2022ofdm}. Apart from the power inefficiency incurred by TPC, its scalability and evolution beyond the \textit{request-receive} communication protocol still remain a big question in MGs. This sets up the foundation for innovations in new co-transfer technologies beyond the state-of-the-art.}\par
 
In this regard, the development of artificial intelligence (AI) has opened up unconventional possibilities to go beyond traditional model-driven norms. {AI-based semantic communication is able to solve the specific challenges of communication in MGs by eliminating key stages, such as signal modulation and communication channels}. Basically, \textit{semantic communication} is a task-oriented approach that does not rely on bit sequences and allows for the exchange of most significant information through various forms of transmission \cite{gunduz2022beyond}. This extends an astonishing opportunity for the power systems communication paradigm, as electrical transients inherently contain valuable information of system dynamics. As a result, the power flows among the lines is hypothesized to be sufficient to distill coordination among different converters in MG. However, a critical requirement is a robust decoder/processor that can effectively \textit{translate} the global communicative signatures from the transients. 
Although the hierarchical evolution using AI in distributed optimization tasks, particularly for secondary control applications \cite{ma2021nonlinear} has been a promising alternative for MGs, {it is essential to exercise caution considering high energy requirements, low efficiency, latency and implementation complexity {of} many neural network (NN) architectures}.
\par

\begin{figure*}[t]\centering
 %   \vspace{-12 pt}
	\includegraphics[width=0.95\linewidth]{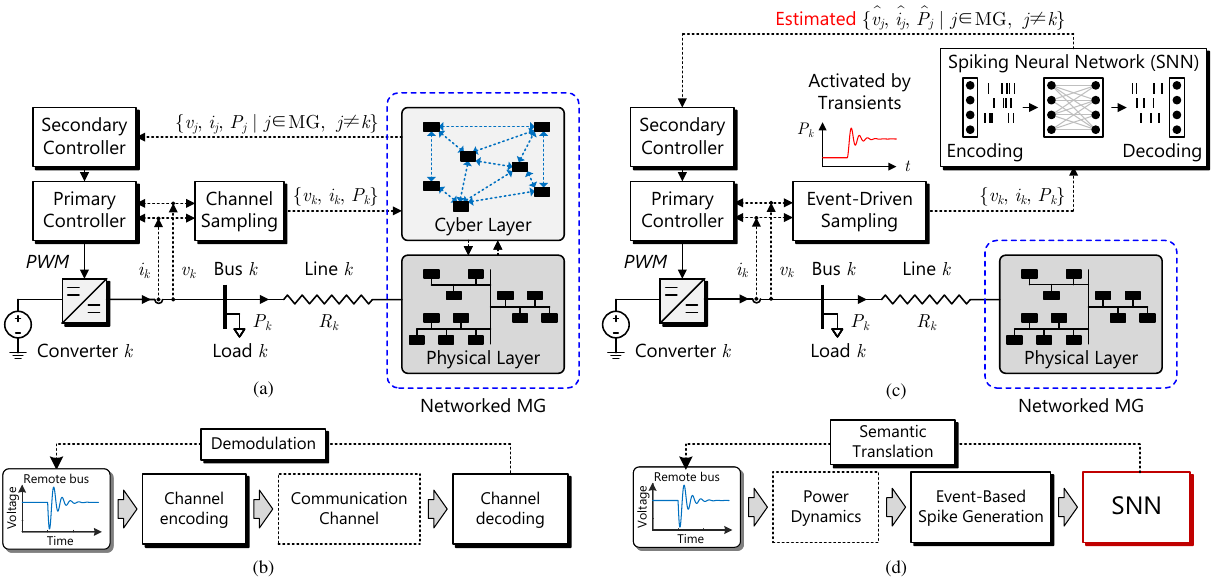}
	\caption{Going beyond traditional communication norms to task-oriented semantic communications in microgrids: (a) conventional cyber-physical control framework, (b) stages in traditional communication, (c) proposed NSC-based coordinated control framework, (d) simplified and reduced stages in NSC.}
    \label{FIG_1}
    \vspace{-6 pt}
\end{figure*}
Inspired by the function of biological brains, we explore by showcasing in DC microgrids {the potential of} \textit{neuromorphic} processors, which exploit Spiking Neural Networks (SNNs) as a low-energy decoder for converters to communicate between each other only using the non-modulated power flows between them. Unlike the next generation artificial neural networks (ANN), which always utilizes real valued samples for learning and inferences, SNN simply compute using binary spikes to excite a change in its corresponding weights by governing local dynamics \cite{jang2019introduction}. As a result, SNN can be trained online in a sparse manner corresponding to any physical disturbances only using local power flow measurements. {Consequently, the proposed inferential mechanism in this paper operates on an entirely different protocol, namely the \textit{publish-subscribe}  architecture {\cite{petar}}.} From an application perspective, dedicated spike-based processors, such as Intel's \textit{Loihi} 2 and IBM's \textit{TrueNorth} offers several advantages:
\begin{itemize}
\item \textit{High energy efficiency}: Since most neurons in a SNN remain idle in the absence of \textit{events}, the spike activity is sparse. This sparsity leads to energy-efficient computations as processing binary spikes \texttt{1} and \texttt{0}, requiring less computational energy compared to the complex operations involving high-precision floating-point numbers used in ANNs \cite{roy2019towards}.
\item \textit{Low latency}: Considering the control application of MGs in this paper, the overall delay of transmitting information packets from one end to another is significantly reduced not only due to information embedding on instantaneous power but also due to the low latency of SNNs itself.
\item \textit{Hardware performance}: Neuromorphic technology is designed with an asynchronous address event representation (AER) architecture, which differs from the rest that still rely on a global clock. This AER architecture aligns well with the asynchronous operation of SNN, enabling low-latency and energy-efficient behavior\cite{deng2020rethinking,nitzsche2022comparison}.  
\end{itemize}

In the recent literature, SNNs have demonstrated successful applications in image classification \cite{vaila2020deep} and wireless communications \cite{chen2023neuromorphic} only using spikes generated from a physical channel/medium to extract contextual information or messages. However, its synthesis process including decoding the spikes to meaningful real-valued information is carried out by powerful neuromorphic sensors in those applications. On the other hand, augmentation of a new spiking-based sensing technology is not a straight-forward mechanism for power systems today, since their operations are only equipped to translate and operate with real-valued measurements. 

This is where we propose an elementary method to generate spikes based on \textit{semantic events} \cite{kirti,kirti2} collected from the existing measurements in microgrids that consequently federates the training of SNN at each node. By doing so, not only the energy consumption can be significantly reduced by collecting binary data instead of floating point numbers, but it will also expedite low-power edge processors. This online training is triggered only when nodal power dynamics exceed a pre-defined event-detection threshold. Finally, these events are encoded and decoded as spikes for remote estimation. As a result, the neuromorphic semantic communication (NSC) allows us to go beyond the current cyber-physical architecture of microgrids {{(as shown in Fig. \ref{FIG_1}(a) and (b)) to a novel intrinsic communication principle (as shown in Fig. \ref{FIG_1}(c) and (d))} for power electronic converters only using the power flows between them. {The comparison of NSC, TPC and traditional cyber layer communication (CLC) are presented in Table \ref{compare}}.

\par
To sum up, the research contributions of this paper are:
\begin{enumerate}
\item A pioneering application of neuromorphic computing in the coordinated control of MGs is proposed. To ensure its suitability for coordinated control of microgrids and prospective extension in larger systems, this paper tailors spiking neural networks (SNNs) as a grid-edge inference technology with detailed steps of implementation. {By eliminating the communication infrastructure,} the proposed philosophy not only removes exogeneous arrival paths for cyber-attackers and other reliability issues in the cyber layer, but also unravels unification of power and information in general {using the \textit{publish-subscribe} protocol} that can be extended for many applications. 
\item Since SNNs operate using the energy-efficient binary spikes in a sparse and online fashion using spike timing dependent plasticity (STDP), we employ \textit{semantic events} translating the dynamic response of each converters with respect to the measured power flows as spikes to carry out online training of NSC. 
 \item Going beyond TPC that allows co-transfer of power and data but are limited by electrically isolated stages, the proposed communication principle is not limited by such stages that physically block the zero-sequence path. Its scalability and flexibility for different voltage levels and system topologies has been investigated in detail. Moreover, it also promises high computational energy efficiency during data processing and learning, that has been bench-marked with respect to binary-activated recurrent neural networks (RNNs) and ANN. 
 \end{enumerate}
 
 {In this paper, DC MGs have been focused to showcase the principle behind the proposed NSC, whereas it is also potentially applicable to AC systems where the data collection methodology may need to be revised.}

\begin{table}[hb]
  \setlength{\tabcolsep}{1.6pt}
    \centering
\caption{{Comparative Evaluation of NSC, TPC and CLC.}}
\label{compa}
    \begin{tabular}{l c c c c} \hline \hline 
         {Methods}&   {\makecell{Prone to \\Delay?}}&  {\makecell{Prone to \\Cyberattack?}}&{\makecell{Prone to \\Channel outage?}} &{\makecell{Prone to \\Inefficiency?}}\\ \hline          {CLC \cite{dong2019stability}, \cite{sahoo2018stealth}}&  {Yes}&  {Yes}&  {Yes} &{No}\\  
         {TPC\cite{angjelichinoski2016multiuser}, \cite{he2020nature}} &  {No}& {No}&  {No} &{Yes}\\ 
 {NSC} & {No}& {No}& {No} &{No}\\\hline\hline 
    \end{tabular}
    \label{compare}
\end{table}
%I wish you the best of success.
\section{Neuromorphic Coordinated Control of Microgrids}
\subsection{Problem formulation and motivation}
The conventional cyber-physical framework of the most reliable infrastructure, i.e., distributed coordination in a DC MG is depicted in Fig. \ref{FIG_1}(a). In this framework, each converter is locally regulated by the primary controller and globally by the secondary controller that relies on exchange of real-time information, including voltage $v_j$ and current $i_j$ from the neighboring nodes $j \in N_k$ in Fig. \ref{FIG_1}(a), where $N_k$ denote the set of neighbors for converter $k$. It is crucial to note that the stability and reliability of MGs heavily rely on the effectiveness of the communication network \cite{bidram2013distributed, sahoo2018stealth}. {However, as presented in Fig. \ref{FIG_1}(b), the conventional communication methods/protocols in MGs include stages such as, modulation, transmission, and demodulation,} which is done in a timely process that introduces the notion of bandwidth and consequently suffers from communication delays \cite{dong2019stability}, cyberattack vulnerability \cite{sahoo2018stealth}, and susceptibility to cyber link outages \cite{wu2022line}. These disturbances can range from few milliseconds to seconds and can negatively impact the dynamic performance and stability of MGs. Although co-transfer technologies such as TPC can eliminate the need for conventional communication infrastructure, it is still susceptible to scalability in high voltage levels (due to mutual inductance between lines causing signal interference) and path blockage beyond galvanic isolation, which can affect communication reliability. {As a result, new intrinsic and scalable means of communication among converters is needed to overcome the said challenges and investigate the ongoing digitalization measures in MGs.} 

This section unravels a novel end-to-end NSC-based coordinated control framework, as depicted in Fig. \ref{FIG_1}(c), which comprises of the SNN as its key component. {Its protocol stages are illustrated in Fig. \ref{FIG_1}(d), where instead of transferring information from one end to the other, remote information can be inferred at each end using the spatio-temporal pattern of the power dynamics corresponding to the physical disturbances. Using event-driven sampling, \textit{meaningful} information is filtered and encoded into spikes for initial weight determination of SNN to be deployed at each bus. Relying on the \textit{publish-subscribe} architecture \cite{petar} depending on a global update that can be translated as any disturbance {in MGs,} SNNs then infer the information of remote buses by measuring power flows. It is worth mentioning that the spike timing dependent plasticity (STDP) feature further allows online training of SNNs only by measuring power flows to adapt its weights accordingly.}

As the proposed framework is significantly different from the traditional cyber-physical arrangements (in Fig. \ref{FIG_1}(a)), we firstly discuss the hierarchical control using conventional communication architecture and then discuss the specific design and implementation steps for integrating neuromorphic based semantic coordination into the control of MGs. Building upon this research in end-to-end low power NSC for MGs, this paper draws inspiration and adapts the proposed communication policy only using power flows as the communication channel for coordinated control of MGs, as shown in Fig. \ref{FIG_1}(c) and (d). 

\subsection{Conventional hierarchical control in MGs}
Fig. \ref{FIG_1}(a) depicts the conventional hierarchical control structure of a DC MG. The primary control uses local measurements to regulate the output voltage based on a $V-I$ droop control strategy. Additionally, the local sampling data is shared with adjacent nodes through the cyber layer, facilitating coordination and monitoring.
\subsubsection{Primary control}
The primary control layer in converter $k$  is implemented as:
\begin{equation}
    v_{ref,\:k}(t)=v_{k}(t)-m_ki_{k}(t)
\end{equation}
where, $m_k$ is the droop gain, which is calculated using $\Delta V_{k}/I_{k_{max}}$ with $\Delta V_{k}$ and $I_{k_{max}}$ being the maximum voltage deviation and maximum output current of converter $k$, respectively. Furthermore, $i_k$ and $v_{ref,k}$ denote the converter output current and voltage reference of converter $k$, respectively. 
\subsubsection{Distributed secondary control}
The secondary controller plays a crucial role in achieving voltage regulation and other energy management schemes. As distributed secondary control offers highly reliable and cost-efficient coordination \cite{sahoo2017distributed}, we consider it as the best candidate for CLC in this paper. With distributed secondary control, the local voltage set point can be expressed as:
\begin{equation}
    v^*_k(t) =  v_{ref,\:k}(t)+\delta v^{\mathrm{I}}_k(t)+\delta v^{\mathrm{II}}_k(t)
\end{equation}
where, $ v_{ref,\:k}$ is the voltage regulation term generated by the primary controller in (1). The first voltage correction term $\delta v^{\mathrm{I}}_k$ is generated by {the voltage observer and further compensated by the secondary PI controller:}
\begin{subequations}
\begin{align}
&\bar{v}_k(t)=v_k(t)+\int_{0}^{t}\sum_{j\in N_k}a_{kj}(\bar{v}_j(t)-\bar{v}_k(t))\mathrm{d}\tau\\
&\delta v^\mathrm{I}_k(t)=k_{pU}(v_{ref}-\bar{v}_k(t))+k_{iU}\int_{0}^{t}(v_{ref}-\bar{v}_k(t))\mathrm{d}\tau
\end{align}
\end{subequations}
{where, $\bar{v}_k(t)$ is the average voltage observed at bus $k$, $N_k$ is the set of nodes that are adjacent to node $k$ in the cyber graph. In a graph with $N$ nodes, each node represent a converter, that are communicating among each other using edges through an associated adjacency matrix, ${\mathbf A_\text{G}} = [{a}_{kj}]\in{R^{N\times{N}}}$, where the communication weight (represented by $a_{kj}$, i.e., from node $j$ to node $k$) is formulated as: $a_{kj} >$ 0, if ($\psi_k$, $\psi_j$) $\in$ $\mathbf{E}$, where $\mathbf{E}$ represents an edge connecting two different nodes, with $\psi_k$ and $\psi_j$ representing a local node and its neighboring node, respectively. If the cyber link connecting $\psi_k$ and $\psi_j$ is absent, $a_{kj}$ = 0. More modeling preliminaries of the cyber graph can be obtained from \cite{10061596}.
Furthermore, the second voltage correction term $\delta v^{\mathrm{II}}_k$ is generated by cooperative regulators for particular objectives. For current sharing, the regulation term is calculated by:}
\begin{subequations}
\begin{align}
    &\lambda_k(t)=\sum_{j\in N_i}a_{kj}(i_j(t)-i_k(t))\\
    &\delta v^{\mathrm{II}}_k(t)=k_{pI} \lambda_k(t)+k_{iI}\int_{0}^{t}\lambda_k(t)\mathrm{d}\tau
    \end{align}
\end{subequations}
{Similarly, for power sharing objective, the regulation term is:} 
\begin{subequations}
\begin{align}
    &\eta_k(t)=\sum_{j\in N_i}a_{kj}(P_j(t)-P_k(t))\\
    &\delta v^{\mathrm{II}}_k(t)=k_{pP}\eta_k(t)+k_{iP}\int_{0}^{t}\eta_k(t)\mathrm{d}\tau
    \end{align}
\end{subequations}
{where, $k_{pI}$, $k_{iI}$, $k_{pP}$, and $k_{iP}$ are the proportional and integral coefficients of PI controllers for current sharing and power sharing, respectively.}\par
{Given that the {physical layer of the MG and cascaded control structure from the primary control loop to the PWM stage remain the same in both {\mbox{Fig. \ref{FIG_1}(a) and (c)}},}} the key distinction of incorporating NSC instead of relying on the traditional CLC lies in how the dynamic measurements from the remote nodes are efficiently predicted to disregard any exogenous arrival paths or unreliable cyber scenarios, which is the main contribution of this paper. In addition to the elimination of a dedicated communication channel that primarily relies on the \textit{request-receive} protocol, we leverage the proposed NSC framework to infer real-time information using power flows by deploying SNN at each bus. We firstly cover the background of the biological neuron modeling for SNNs and then discuss the offline initial weight determination of SNN at each bus in the upcoming subsections.
\begin{figure}[!thb]\centering

	\includegraphics[width=\linewidth]{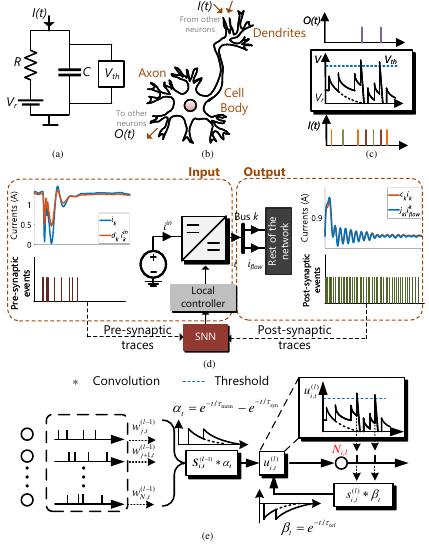}
	\caption{{Initial weight determination of SNN at bus $k$ -- Pictorial depiction of a (b) RC circuit and its structural duality with (b) biological neuron, where the input current $I(t)$ acts as an excitation signal into both the circuit and the neuron -- its excitation dynamics are then translated into \textit{meaningful} output spikes for the learning of NSC using a (c) voltage threshold $V_{th}$ based criteria, (d) Similar to the biological neurons in (b), a converter having disturbances in its input and output can consequently correspond to pre-synaptic and post-synaptic events, respectively, (e) Information-theoretic learning based communication using spiking neural networks (SNNs) and the spike response model (SRM) for simulation of a LIF neuron.}
    %: spike response model (SRM), coding methods and backpropagation algorithm.
    }
    \label{FIG_42}
\end{figure}
{\subsection{Fundamentals of SNN operation}
Instead of the summation functions in multilayer perceptron networks for ANN, SNN employs neuronal dynamics that rely on the integration process with a framework that triggers action potential only above a critical voltage \cite{snnbas}. As shown in Fig. \ref{FIG_42}(b), a biological neuron is basically excited by a current $I(t)$ as a pulse input coming from the nearby neurons into the axon of the cell body. Similar to the electrical properties in a RC circuit (see Fig. \ref{FIG_42}(a)) with a cell voltage $V_r$, the current $I(t)$ will only flow given that the capacitor voltage $V_{th}$ is more than $V_r$. Mathematically, this can be represented by:
\begin{equation}
    I(t) = \frac{V_{th}(t) - V_r}{R} + C\frac{\mathrm{d}V_{th}}{\mathrm{d}t}
\end{equation}
Considering the time constant $\tau_m$ = $RC$ as the \textit{leaky integrator}, we get:
\begin{equation}
    \tau_m \frac{\mathrm{d}V_{th}}{\mathrm{d}t} = -[V_{th}(t) - V_r] + RI(t)
\end{equation}
Translating the electrical specifications into the neuron cell, we can then refer $V_{th}$ to be the membrane potential and $\tau_m$ to be the membrane time constant of the neuron. Due to the notion of leakage of charge with a time constant $\tau_m$ and (7), this neuron model is commonly called as the leaky-fire and integrate (LIF) model. Approximating the LIF neuron dynamics in Fig. \ref{FIG_42}(b) using the Euler method \cite{snnbas}, we get:
\begin{equation}
    \tau_m \frac{\mathrm{d}V_{mem}}{\mathrm{d}t} = -(V_{mem} - V_{th}) + \frac{I_s}{g}
\end{equation}
where, $V_{mem}$ is the membrane potential, $g$ is the leaky conductance, $V_{th}$ is the threshold, $I_s$ is the synaptic current and $\tau_m$ is the membrane time constant.

Considering the dynamics in (8) and selection of a corresponding event-triggering criteria \cite{etr}, the input spikes can then be determined corresponding to rising or falling edges in $I(t)$. As shown in Fig. \ref{FIG_42}(c), the input spikes that manage to invoke beyond the membrane potential $V_{th}$ result into a selective number of output spikes.

Extending the single neuron structure to a bidirectional DC/DC converter in Fig. \ref{FIG_42}(d), the current excitation can either emanate from the input, such as intermittent generation from renewable energy sources, or the output, such as load change or tie-line outage. Since the current flow is in both directions as compared to the case in Fig. \ref{FIG_42}(b), we decipher both input as well as output dynamics to achieve accuracy in the estimation of information at remote buses. Using multiple data set-points corresponding to different operational scenarios, semantic data collection is performed to assign the initial weights of SNN, which will be discussed later. 
}
\subsection{Network model of SNN}
Before discussing the weight initialization strategy for SNNs to be deployed at each bus, we firstly uncover the underlying theory behind the entire network to be modeled. The neuron model in Fig. \ref{FIG_42}(e) is the spike response model (SRM). It is a widely recognized model that effectively represents the characteristics of biological neurons while retaining simplicity, which makes it well-suited for the intended application of the NSC-based coordinated control framework in MGs \cite{chen2023neuromorphic}.\par
In Fig. \ref{FIG_42}(e), the neuron $N_{i,l}$ is connected with its preceding neurons by different synaptic weights $\{w_{j,l-1}\|\ j\in \mathrm{Layer}\: (l-1)\}$. The preceding neurons can transmit binary spikes to  $N_{i,l}$. The membrane potential $ u_{i,l}$ of $N_{i,l}$ represents an analog state to describe the contribution of the spikes. When  $N_{i,l}$ receives a spike, $ u_{i,l}$ momentarily increases and decays exponentially over time. This behavior is described as the second-order synaptic filter, $\alpha_t=e^{-t/\tau_{m}}-e^{-t/\tau_{syn}} $. Referring to (8), the decay process is modeled as the first-order feedback filter $\beta=-e^{-t/\tau_{ref}}$ with finite positive constants $\tau_{m},\tau_{syn},\tau_{ref}$. The membrane potential $u_{i,l}$ is computed as the sum of the filtered contributions from incoming spikes and the neuron's own past outputs as follows:
\begin{equation}
 u^{(l)}_{i,t}=\sum_{j=1}^{N}w^{(l-1)}_{j}\cdot(\alpha_t*s^{(l-1)}_{j,t})+\beta_t*s^{(l)}_{i,t}
\end{equation}
where, $\ast$ denotes the convolution operator, and $s^{(l-1)}_{j,t}$ represents the spike from layer $(l-1)$. The spike $s^{(l)}_{i,t}$ is generated by neuron $N_{i,l}$ at time step $t$ when its membrane potential $u^{(l)}_{i,l}$ surpasses the threshold $U_{thr}$, using:
\begin{equation}
    s^{(l)}_{i,t} = H(u^{(l)}_{i,t}-U_{thr})
\end{equation}
where, $H(\cdot)$ is the Heaviside step function, given by: 
\begin{equation}
H(x)=\left\{
\begin{array}{rcl}
1 & & {x > 0}\\
0 & & {x \leq 0}
\end{array} \right.
\end{equation}
{\textit{\textbf{Remark I:} Unmodulated power flows with pub-sub communication protocol inherently imbibe communicative features due to the spatio-temporal patterns of voltage $V$ in power electronic grids, where $V$ = $V_0 e^{-a}$ such that $a$ = $\{t, x\}$ corresponding to time and space, respectively. As SNNs are highly capable of synthesizing these spatio-temporal patterns in an energy-efficient and online fashion, we underline our proposal with SNN as an energy-efficient grid-edge intelligence tool for coordination and health monitoring of future power systems.}}
\subsection{Data collection and offline design of SNN}
%Considering the specific dynamic characteristics of MGs depicted in Fig. \ref{FIG_1}(b), the encoding stage of SNN is bypassed as the power dynamics resulting from events can inherently convey global information to each converter. Instead of a dedicated communication channel, the power flows via tie-lines itself is exploited to govern power dynamics at remote ends, such that the communicative signatures can be extracted at each remote nodes. 
%Additionally, a decoder is implemented at the receiver to interpret the received information from remote ends, which are influenced by power dynamics, into the intended target information.\par
%This is made possible by an unconventional methodology, i.e., the proposed NSC framework in Fig. 1(d) that predicts most significant remote information using spikes based events. 
{It is worth notifying that \textit{events} in this paper denote the physical disturbances in the system.} Such disturbances either in the input or output of DC/DC converter in Fig. \ref{FIG_42}(d) are essentially formalized by the power dynamics turned events locally at each bus with the power lines being the propagating medium itself. Hence, {the input dynamics in Fig. \ref{FIG_42}(d)} of DC/DC converter $k$  associated with the capacitor current $i^C_k$ and inductor voltage $v^L_k$ can be given by:
\begin{eqnarray}
\begin{cases}
i^C_k (t) = C_k\frac{\mathrm{d}v_k}{\mathrm{d}t} = i_k (t) - d_k i^{in}_k (t)\\
v^L_k (t) = L_k\frac{\mathrm{d}i_k}{\mathrm{d}t} = v_k (t) - d_k v^{in}_k (t)
\end{cases}
\end{eqnarray}
where, $d_k$ denotes the voltage amplification ratio between the input and output end of DC/DC converter $k$, $i^{in}_k$ and $v^{in}_k$ denote the input current and voltage of converter $k$, respectively. Moreover, $L_k$ and $C_k$ denote the inductor and capacitor in the DC/DC converter $k$, respectively. Finally, the corresponding error signals for both currents and voltages to formalize/trigger the input \text{events} in $k^{th}$ converter {(see Fig. \ref{FIG_42}(d))} are given by:
\begin{eqnarray}
    \begin{cases}
    \Omega_i (t) = v^L_k (t) - e^i_k (t) \\
    \Omega_v (t) = i^C_k (t) - e^v_k (t)
    \end{cases}
\end{eqnarray}
where, $e^i_k(t) = i^{in}_{ref,k} (t) - i^{in}_k (t)$ and  $e^v_k(t) = v_{ref,k} (t) - v_k (t)$. Since the error values $e^V_k$ and $e^i_k$ have a low time constant with respect to the given switching frequency as compared to the ones using CLC, the physical system semantics is used to extrapolate the most significant data to be collected for each converter. Not only this data collection principle allow collection of qualitative data, it ensures the most significant information to be distilled for effective training of the provisional SNN. 
\algrenewcommand\algorithmicrequire{\textbf{Input:}}
\begin{algorithm}[b]
\caption{{Input} event translation to spikes.}
\label{alg:eventcapturing}
\begin{algorithmic}[1]
\Require $m^\mathrm{th}$ sample of voltage $v[m]$, current $i[m]$\par
         Thresholds of voltage and current variance to trigger and hold an event $\sigma^{V}_{th}$, $\sigma^{I}_{th}$\par
\State $V^{ben}[0] \gets 0$, $I^{ben}[0] \gets 0$
\State Verify $\Omega_{v}$ and $\Omega_{i}$ using (16)
\If {$\Omega_{v}\:\texttt{OR}\:\Omega_{i}$}
    \State \texttt{event}[$m$] starts
%    \State $k = 0$ and recount samples from 0
    \Repeat
        \State SNN is activated
        \State // Update the benchmarks
        \State Update $V_{ben}[m]$: $V_{ben}$=1
        \State Update $I_{ben}[m]$: $I_{ben}$=1
        \State $m \gets m+1$
        \State Re-verify $\Omega_{v}$ and $\Omega_{i}$ by Eq. (16)
    \Until {$\texttt{NOT}\:(\Omega_{v}\:\texttt{OR}\:\Omega_{i})$}
    \State \texttt{event}[$M$] ends
    \State Hold $V_{ben}[M]$ and $I_{ben}[M]$ as the steady-state values
\EndIf
\end{algorithmic}
\end{algorithm}

To scale from the local to global {events}, remote estimation using the {output dynamics of all the converters in Fig. \ref{FIG_42}(d)} can be further distilled into a vector representation:
\begin{equation}
    \mathbf{C}\mathbf{\dot{V}}(t) = \mathbf{JI_{flow}}(t) - \mathbf{d i^{in}} (t)
\end{equation}
where, $\mathbf{J}$ is a row matrix with binary values, such that $j_{kl}$ will be 1 only if there is a direct physical connection between converter $k$ and $l$, or otherwise. Moreover, $\mathbf{I_{flow}}$ is a column matrix that comprises the tie-line flow currents into the connected lines resulting out of the output current $i_k$, such that $i_k = \sum i_{flow}$. {Furthermore, $\mathbf{C}, \mathbf{V}, \mathbf{d}$ and $\mathbf{i^{in}}$ denote $k \times k$ diagonal matrices for $C_k$, $v_k$, $d_k$ and $i^{in}_k$}. {The formalization of output events can then be carried out using:
\begin{equation}
    \Omega_{o}(t) = C_k \dot{v}_k(t) - \dot{I}^k_{flow}(t)
\end{equation}}
{Using the spatio-temporal pattern exploration hypothesis in Remark I}, for a given network admittance matrix, each output current has a given distribution of intrinsic communication signatures in the form of $i_{flow}$, that can be estimated using the following data collection process. Finally, sparse sampling and data collection is formalized only the semantic event detection criteria in (13) and (15) exceed a given threshold:
\begin{equation}
    || \Omega_{v}(t) || > \sigma^{V}_{th}, \ \ \ ||\Omega_{i}(t)|| > \sigma^{I}_{th}, \ \ \ ||\Omega_{o}(t)|| > \sigma^{o}_{th}
\end{equation}
where, $\sigma^{V}_{th}$, $\sigma^{I}_{th}$ are the thresholds for input events in (13) and $\sigma^o_{th}$ denote the threshold for the output event in (15), respectively. To achieve good resiliency against noise, a state-dependent threshold can be used \cite{noise}.

The {input} events are then translated into spikes using \texttt{Algorithm 1} such that SNN exploits the local measurements and their dynamics to estimate remote measurements. {The output spikes can also be generated by following \texttt{Algorithm 1} for the dynamics in (14) and triggering criteria in (15).} Finally using the NSC-based control method, the remote voltage and current in (3a), (4a) and (5a) are replaced with the estimated values $\hat{\circ}$, as follows:
\begin{subequations}
\begin{align}
&\bar{v}_k(t)=v_k(t)+\int_{0}^{t}\sum_{j\in N_{flow}}(\hat{\bar{v}}_j(\tau)-\bar{v}_k(\tau))\mathrm{d}\tau\\
&\lambda_k(t)=\sum_{j\in N_{flow}}(\hat{i}_j(t)-i_k(t))\\
&\eta_k(t)=\sum_{j\in N_{flow}}(\hat{P}_j(t)-P_k(t))
\end{align}
\end{subequations}
where, $\hat{P}_j(t)$ is obtained by {$\hat{P}_j(t)=\hat{i}_j(t) \hat{v}_j(t)$ and $N_{flow}$ is the set of nodes that are adjacent to converter $k$ in the physical tie-line admittance network graph. Having discussed the {offline preliminary design of SNN at each node, we now discuss its online training based on the spike timing dependent plasticity (STDP) feature in the next section.}

\section{Spike Timing Dependent Plasticity}

\subsection{Online weight adaptation of SNN}
       After obtaining the preliminary offline design of SNN, the dynamic weight adaptation corresponding to different transients/events is carried out using \textit{Spike timing dependent plasticity} (STDP) \cite{stdp}. STDP is a neurobiological concept that describes how the strength of a synapse in Fig. \ref{FIG_42}(b), which is the connection between two neurons, can be modified based on the precise timing of the spikes from inputs and outputs or the action potentials in these neurons. It is a fundamental mechanism that underlies learning and memory in the brain, which makes it a biology-plausible training method and allows online adaptation of the weights of SNN. The mathematical formulation behind the STDP based weight update policy is as follows:
    \begin{equation}
        \Delta W=\left\{
        \begin{array}{rcl}
         A_{+}e^{(t_{pre}-t_{post})/\tau_{+}} & & {(t_{post} > t_{pre})}\\
         -A_{-}e^{-(t_{pre}-t_{post})/\tau_{-}} & & {(t_{post} < t_{pre})}
         \end{array} \right. 
    \end{equation}
    
    where,  $\Delta W$  denotes the change in the synaptic weight of SNN,  $A_{+}$  and  $A_{-}$  determine the maximum amount of synaptic modification (which occurs when the timing difference between pre-synaptic and post-synaptic spikes in Fig. \ref{FIG_42}(d) is close to zero),  $\tau_{+}$  and  $\tau_{-}$ determine the ranges of pre-to-postsynaptic inter-spike intervals over which synaptic strengthening or weakening occurs. As illustrated in Fig. \ref{STDP}(a), $t_{pre}$ and $t_{post}$ are the timings of the pre-synaptic and post-synaptic spikes, that correspond to the input and output disturbances of the DC/DC converter in Fig. \ref{STDP}(b), respectively. As illustrated in Fig. \ref{STDP}(a), if the postsynaptic neuron spikes arrive after the pre-synaptic neuron, $\Delta W>0$, {or otherwise, $\Delta W<0$}. The weight update policy and its biological plausibility is explained by the Hebbian Principle, that is often summarized as ``neurons that fire together wire together". If a pre-synaptic neuron fires just before a postsynaptic neuron, the connection between them is strengthened, often known as long-term potentiation (LTP) of the synapse \cite{rich}. Otherwise, the connection is weakened, often known as long-term depression (LTD) of the same synapse.\par
%    We have to clarify here that STDP is not limited by the encoding and decoding methods [R17]. Since it only focuses on the individual timing of the pre-spikes and post-spikes. That means no matter how many spikes are involved in the synapse updating process, we can always analyze them individually, so both rate encoding and temporal encoding are compatible with STDP. Here is the analysis process:
Based on the abovementioned definitions, it is vital to analyze the time difference between the pre-synaptic and post-synaptic spikes, given by: 
\begin{equation}
        \Delta t=t_{pre}-t_{post}
    \end{equation}
From a DC/DC converter perspective, this would imply that the time difference $\Delta t$ could either arise depending on the physical disturbances that is either in the input or output stage, as illustrated in Fig. \ref{STDP}(b).

       \subsection{Calculation of $\sum^Z_{i=1}e^{\Delta t_i}$ and $\sum^Z_{i=1}e^{-\Delta t_i}$}
       Keeping track of the pre-and postsynaptic spikes for the number of neurons from 1 to $Z$, a neuron will receive numerous pre-synaptic spike inputs that needs to be processed simultaneously by the SNN. The post-synaptic neuron can be defined by:
\begin{equation}
    \tau_{-}\frac{\mathrm{d}L}{\mathrm{d}t}=-Q
\end{equation}
and for every postsynaptic neuron spikes, it is updated using:
\begin{equation}
    Q(t)=Q(t)-A_-
    \label{eq_15}
\end{equation}
In this manner, $Q(t)$  tracks the number of postsynaptic spikes over the given timescale  $\tau_{-}$. Similarly, for each pre-synaptic neuron, we define:
\begin{equation}
    \tau_{+}\frac{\mathrm{d}R}{\mathrm{d}t}=-S
\end{equation}
and for every spike on the pre-synaptic neuron, it is updated using:
\begin{equation}
    S(t)=S(t)+A_+
    \label{eq_17}
\end{equation}
It is worth notifying that the variables $Q(t)$ and $S(t)$ are quite similar to the notion of synaptic conductance $g(t)$ in (8), except that they are particularly defined for spike timings on a much longer timescale. Based on the illustration of $\Delta W$ in Fig. \ref{STDP}(a), $S(t)$ is inherently negative that is used to induce LTD ($\sum^N_{i=1}e^{-\Delta t_i}$) and $Q(t)$ is always positive used to induce LTP ($\sum^N_{i=1}e^{\Delta t_i}$). The reason behind the negative and positive signs of $S(t)$ and $Q(t)$ are because they are updated by  $A_-$ and $A_+$, respectively. As illustrated in Fig. \ref{STDP}(b), the weight increment/decrement at $t_{pre}$ and $t_{post}$ can be given by:
\begin{eqnarray}
    \Delta W(t_{pre}) = S(t_{pre})W(t_{pre})\\
    \Delta W(t_{post}) = Q(t_{post})W(t_{post})
\end{eqnarray}
To implement STDP in SNN, based on the pre-synaptic and post-synaptic timing, update of the variables $Q(t)$  and $S(t)$ vary their synaptic conductance. With the peak synaptic conductance $g_i$ for a synapse $i$ bounded between [0, $g_{max}$], it is modified accordingly based on either LTP or LTD condition using:

\begin{equation}
    \bar{g}_i=\begin{cases}
        \bar{g}_i+Q(t)\bar{g}_{max} \ \textit{if} \ \textbf{LTP}\\
        \bar{g}_i+S(t)\bar{g}_{max} \ \textit{else}
    \end{cases}
    \label{eq_18}
\end{equation}
Its update corresponding to the LTD or LTP conditions in Fig. \ref{STDP}(b) can be seen in Fig. \ref{STDP}(c), that is used to update the synaptic conductances of the SNN allowing online adaptation corresponding to any transient in the local measurements. As illustrated in Fig. \ref{STDP}(c), the total excitatory synaptic conductance $g_E(t)$ for $Z$ pre-synaptic neurons is given by:
\begin{equation}
    g_E(t)=\sum^Z_{i=1}g_i(t)
\end{equation}
 \begin{figure}[t]
     \centering
     \includegraphics[width=\linewidth]{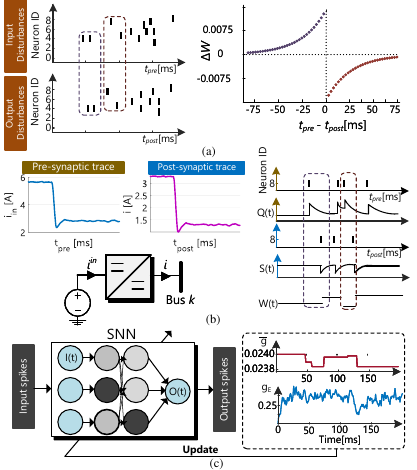}
     \caption{ Online weight update policy of SNN based on the spike-dependent timing plasticity (STDP): (a) Long-term potentiation (LTP) and depression (LTD) based on the excitation of the disturbance observed initially either in the input or the output of the DC/DC converter leading to the trajectory of weight update as per (18), (b) Update of the variables $Q(t)$ and $S(t)$ to formalize the SNN weights as per (24)-(25), (c) update of the synaptic conductance $g_i$ (neuronal) and $g_E$ (total) corresponding to the LTP and LTD events.}
     \label{STDP}
 \end{figure}
 In this way, the online weight update policy subject to selective spikes update the SNN network weights for every physical disturbance in the MG.

\subsection{Design of SNN for MGs}
To optimize the application of SNN in control for MGs, it is necessary to tailor the design of the SNN specifically for this purpose.

To capture the non-linear dynamics accurately, the derivatives of the typical measurements, bus voltages and output currents of each converter are considered as inputs for their respective SNN. To enhance SNN performance, we employ an input layer of 256 neurons. This allows for a more comprehensive representation of the input signals and facilitates the extraction of relevant features. The number of neurons in the output layer depends on the number of output signals required. In this study, each source needs to receive remote voltage and current information of the remaining buses. For instance, in a three-bus system in a symmetric ring topology, each source requires voltage and current data from the rest of the two sources. For an asymmetric radial topology, the number of inputs will vary based on the physical admittance matrix corresponding to each bus. This means that the SNN should have at least 4 output neurons to provide this information. The number of hidden layers and neurons in each hidden layer has been selected after adjudging a significant balance between accuracy and efficiency. The parameters used for designing SNNs in this paper can be found in Appendix.

\section{Results and Discussions}
\subsection{Simulation results}
To assess the effectiveness and versatility of the proposed NSC-based coordinated control across various voltage levels, topologies, control objectives, and its robustness against the SST and line outages, we consider four test cases (shown in Fig. \ref{FIG_5}) and an IEEE 14 bus system in MATLAB/Simulink environment. The training process was carried out using Python, and the model was saved for integration into Simulink for implementation of NSC based control. The description of each disturbance has been categorized into time windows, also termed as ``stages", can be found in Table II.
The system parameters specific to each scenario, starting from Case I to Case V, are provided in Table III. For all the case studies, it should be noted that estimated values from SNN are depicted as $\hat{\circ}$, whereas the measured values as $\circ$ for voltage, current and power.
For simplicity, the dataset only includes disturbances, such as load step changes and line outages for a preliminary investigation of secondary controller. More dynamic disturbances will be considered as a future scope of work to investigate the sensitivity and stability of its performance using a high-performance SNN.\par
\begin{table}[tbh]
  \setlength{\tabcolsep}{1.6pt}
    \centering
\caption{{Description of the Stages in Simulation Studies.}}
\label{tab:my_label}
    \begin{tabular}{l c c c c } \hline\hline 
         {Cases}& \makecell{Stage I \\($t_{1\mbox{-}2}$)}&   \makecell{Stage II\\($t_{2\mbox{-}3}$)}&   \makecell{Stage III\\($t_{3\mbox{-}4}$)}&  \makecell{Stage IV\\($t_4$-end)}\\  \hline 
         Case I&  Load increase&  Load decrease&  Line outage& --\\ 
         Case II&  Load increase&  Line outage&  Line outage& Load decrease\\ 
         Case III&  Load increase&  Load decrease&  Line outage& --\\ 
         Case IV&  Load increase&  Load decrease&  Line outage& Load decrease\\ 
         Case V&  PV power decrease&  PV power increase& Load decrease& --\\
         \hline\hline
    \end{tabular}
\end{table}

\begin{figure*}[!t]\centering
	\includegraphics[width=0.86\linewidth]{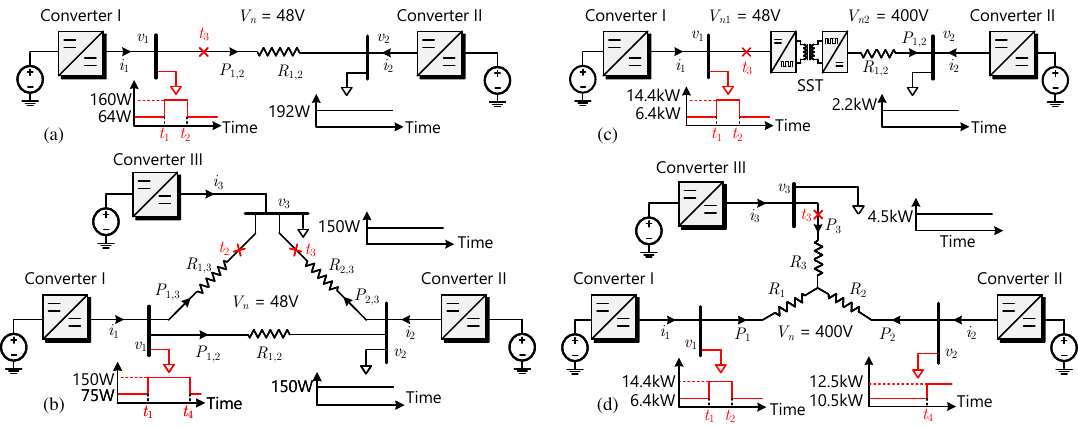}
	\caption{Different topologies for verification: (a) Case I: Two-bus DC microgrid, (b) Case II: Three-bus DC microgrid in ring topology, (c) Case III: Two-bus DC microgrid with an intermediate solid-state transformer, (d) Case IV: Three-bus DC microgrid in star topology.}
    \label{FIG_5}
    \vspace{-12 pt}
\end{figure*}

\begin{figure*}[!t]\centering
	\includegraphics[width=0.95\linewidth]{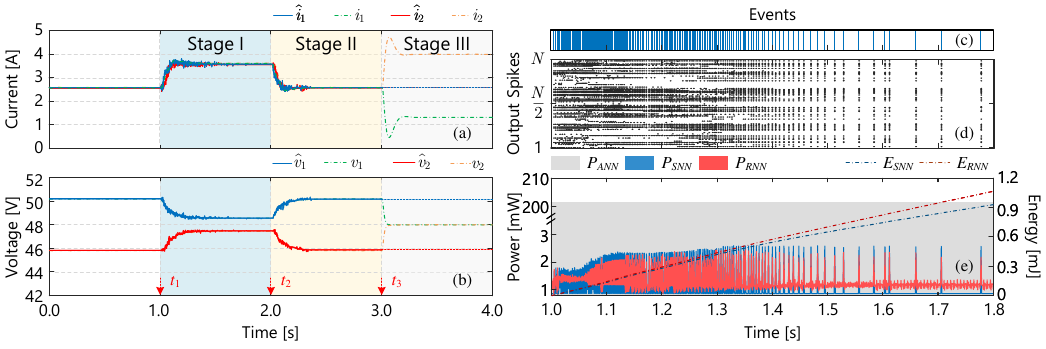}
	\caption{{Case I: (a) SNN-estimated and sampled currents, (b) SNN-estimated and sampled voltages, (c) captured events, (d) output spikes of SNN, and (e) comparison of SNN against ANN and RNN on power and accumulated energy consumption during a load transient.}}
 %comparative evaluation between ANN and SNN of the normalized data numbers in (23).
    \label{FIG_6}
\end{figure*}
\begin{figure*}[!t]\centering
	\includegraphics[width=0.95\linewidth]{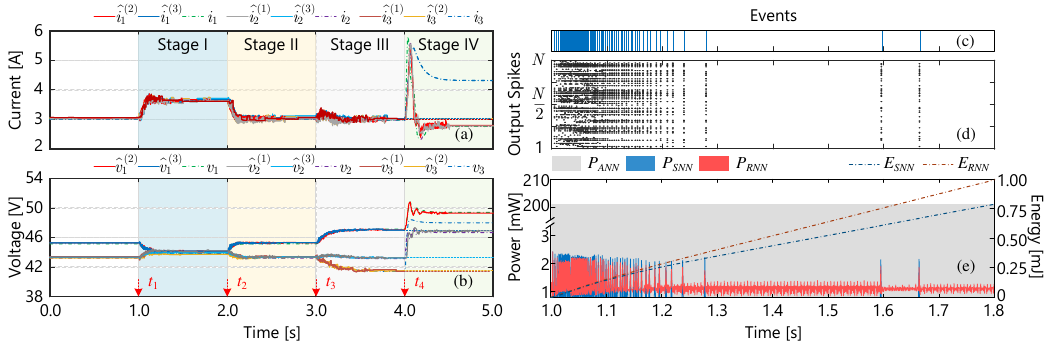}
	\caption{{Case II: (a) SNN-estimated and sampled currents, (b) SNN-estimated and sampled voltages, (c) captured events, (d) output spikes of SNN, and (e) comparison of SNN against ANN and RNN on power and accumulated energy consumption during a load transient.}}
 %comparative evaluation between ANN and SNN of the normalized data numbers in (23)
    \label{FIG_7}
    \vspace{-12 pt}
\end{figure*}

\begin{figure*}
\centering
    \begin{minipage}[b]{.48\textwidth}
    \includegraphics[width = 0.9\columnwidth]{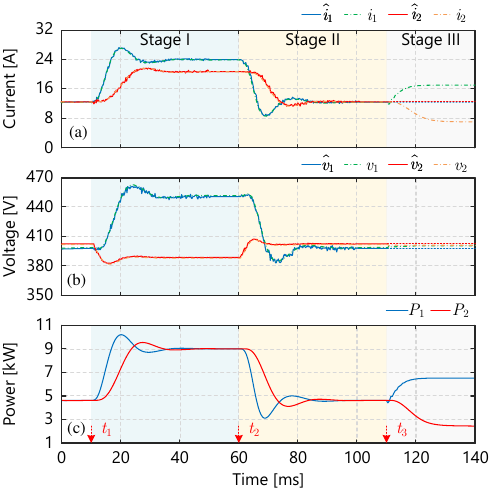}
    \caption{Case III: {(a) comparison of estimated and measured voltages, (b) currents, and (c) output powers}.}
    \label{FIG_9}
    \vspace{-12 pt}
\end{minipage}\qquad
    \begin{minipage}[b]{.48\textwidth}
    \includegraphics[width = 0.9\columnwidth]{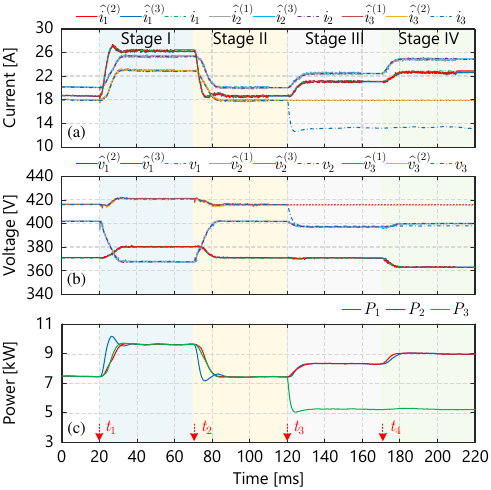}
    \caption{Case IV: {(a) comparison of estimated and measured voltages, (b) currents, and (c) output powers}.}
    \label{FIG_10}
    \vspace{-12 pt}
\end{minipage}
\end{figure*}

\subsubsection{Case I: Two-Bus DC microgrid}

The secondary control objective of Case I is proportionate current sharing.
At time $t_1$ in Fig. 5(a), a step increase in the load from 64 W to 160 W occurs in stage I. As a result of NSC-based coordinated control, $i_1$ and $i_2$ rise simultaneously and reach the same steady value, thereby fulfilling the secondary control objective. This is made possible by the NSC-based coordinated control in (17b) that provides the estimated values $\hat{i}_2$ and $\hat{v}_2$ after decoding from SNN during dynamic processes shown in Fig. 5(b). Furthermore, $v_1$ decreases due to the $V-I$ droop control described in (1), while $v_2$ increases to provide additional power from converter II for current sharing. Similarly at the instant $t_2$, when the system load decreases to 64 W, both $i_1$ and $i_2$ decrease together and reach the same steady value, confirming the effectiveness of the current sharing control. \par
At time $t_3$ (stage III), a line outage occurs between the converters, resulting in islanded operation of both converters. Consequently, the entire secondary control loop, including the SNN represented by the dotted lines, becomes inactive.\par
In Fig. \ref{FIG_6}(c), we present the spikes corresponding to $N$ = 256 neurons between the output layer and the hidden layer during the time interval [1, 1.4] s. As depicted in Fig. \ref{FIG_6}(a) and (b), during [1, 1.2] s, the currents and voltages undergo dynamic changes, and after 1.2 s, their variations become less pronounced. Accordingly, the spikes in Fig. \ref{FIG_6}(c) also exhibit more dynamic behavior during the initial phase and become relatively stable as the steady state is reached.\par
\begin{table}[t]
    \centering
    \caption{{System Parameters of different Simulation cases.}}
    \label{pa}
    \begin{tabular}{ c c c c} \hline \hline 
        &  Parameter&  Symbol&Specification\\ \hline 
        &  Rated voltage& $ V_n$&48 V\\ 
        Case I &  Rated power&  $P_1$=$P_2$&300 W\\ 
        &  Line resistance& $ R_{12}$&1.5 \textOmega\\ 
        &  Line inductance& $ L_{12}$&50 \textmu H\\ \hline 
        &  Rated voltage&  $V_n$&400 V\\  
        Case II&  Rated power&  $P_1$=$P_2$=$P_3$&10 kW\\ 
        &  Line resistance&  $R_1$, $R_2$, $R_3$&1.5 \textOmega, 1.8 \textOmega, 2 \textOmega\\ 
        &  Line inductance& $L_1$, $L_2$, $L_3$&50 \textmu H, 60 \textmu H, 66 \textmu H\\ \hline
        &  Rated voltage&  $ V_n$&48 V/400 V\\ 
        Case III&  Rated power& $P_1$=$P_2$&10 kW\\ 
        &  Line resistance&  $ R_{12}$&3 \textOmega\\
        &  Line inductance&  $L_{12}$&1.5 mH\\ \hline 
        &  Rated voltage& $ V_n$&400 V\\ 
        Case IV&  Rated power&  $P_1$=$P_2$=$P_3$&10 kW\\
        &  Line resistance& $ R_1$, $R_2$, $R_3$&2.4 \textOmega, 1.2 \textOmega, 2.8 \textOmega\\ 
        &  Line inductance& $L_1$, $L_2$, $L_3$&1 mH, 0.5 mH, 0.75 mH\\ \hline 
        &  Rated voltage&  $ V_n$&400 V\\ 
        Case V& Rated power&  $P_1$=$P_2$=$P_3$&15 kW\\
        &  ES voltage&  $V_\mathrm{ES1\mbox{-}4}$&96 V\\ 
        &  MPPT voltage& $V_\mathrm{MPPT}$ &245.6 V\\
        \hline\hline 
    \end{tabular}

\end{table}

The spikes from various neurons are depicted in Fig. \ref{FIG_6}(d), generated in correspondence with the events obtained using (16) illustrated in Fig. \ref{FIG_6}(c). The computational energy consumption, shown in Fig. \ref{FIG_6}(e), dynamically changes in tandem with the events. The energy efficiency of the operation of SNN is calculated in comparison with that of a binary recurrent neural network (RNN) based on a hard sigmoid activation function \cite{NIPS2015_3e15cc11} and ANN. The energy consumption of the neural network includes the consumption of synaptic operations and neuron operations, which is defined by the well-known energy analysis tool \textit{KerasSpiking} for neural networks. In the synaptic operations, the output from the front layer is multiplied by the weight of each synapse. It should be noted that the overall energy consumption is not only affected by the abovementioned software operations but also by the hardware accelerators and its specifications. Neuromorphic computing with fully parallel crossbar array based processors with an inherent asynchronous address event representation (AER) architecture run only in the spiking domain for its inputs/outputs, that lowers the energy consumption. On the other hand, the clock based GPU processors for ANN only accept floating points as the data format, which increases the operations and computational power. Furthermore, SNN also uses the notion of leakage, which not only restrict its operation during events, but also limit the number of neurons in operation due to the Hebbian learning principle. However, ANN and binary-activated RNN are always in operation, which thereby increases the computation power. Hence, the energy efficiency in this paper only accounts the data format and asynchronous event based operation as the comparative aspects based on the benchmarking of SNN against ANN and binary-activated RNN.\par
The number of accumulations (ACC) and multiplication accumulations (MAC) in synaptic and neuron operations are the fundamental energy consumption units that need to be considered for energy consumption \cite{lemaire2022analytical}. The number of ACC and MAC for SNN, binary RNN and ANN is summarized in Table \ref{AC_MC}, where  $N_{ACC}^{ANN}$, $N_{MAC}^{ANN}$, $N_{ACC}^{RNN}$, $N_{MAC}^{RNN}$, $N_{ACC}^{SNN}$ and $N_{MAC}^{SNN}$ are the number of ACC and MAC of the three said NNs. $ N_{in}$, $N_{out}$ are the number of neurons of the pre-synaptic and post-synaptic layer. $ N_{spk}$ in the number of spikes in SNN and binary RNN. Then the energy consumption can be estimated by (\ref{energya})-(\ref{energyc}).
\begin{table}[b]
    \centering
\caption{{ ACCs and MACs in SNN and ANN.}}
\label{AC_MC}
    \begin{tabular}{ccccc} \hline  \hline 
         &  Synaptic operations& \multicolumn{3}{c}{Neuron update }\\\hline
        $N_{ACC}^{ANN}$  &  0& \multicolumn{3}{c}{$ 2N_{out}$  }\\ 
        $N_{MAC}^{ANN}$  &  $N_{in}\times N_{out}$&\multicolumn{3}{c}{$3N_{out}$  }\\ 
        $N_{ACC}^{RNN}$  &  $N_{spk}\times N_{out}$&  \multicolumn{2}{c}{$ N_{out}$ (active)} &0 (idle)\\ 
        $N_{MAC}^{RNN}$  &  0&  \multicolumn{3}{c}{$ N_{out}$}\\ 
        $N_{ACC}^{SNN}$  & $N_{spk}\times N_{out}$ & \multicolumn{2}{c}{$2N_{out}$ (active)} &0 (idle)\\  
        $N_{MAC}^{SNN}$  & 0 &\multicolumn{3}{c}{$N_{out}$}\\ \hline\hline

    \end{tabular}
\end{table}

\begin{table}[b]
\vspace{-12 pt}
	\centering
	\caption{Comparative Evaluation of the Computational Energy.}
	\label{energycon}  
	\begin{tabular}{ccc}
		\hline\hline\noalign{\smallskip}	
		&Case I ($t_{1\mbox{-}2}$) & Case II ($t_{1\mbox{-}2}$)  \\
    \noalign{\smallskip}\hline\noalign{\smallskip}
		$E_{SNN}$ (mJ) & 0.921&0.788\\
        $E_{RNN}$ (mJ) & 1.057&0.991\\
        $E_{ANN}$ (mJ) & 168.47&168.47\\
\noalign{\smallskip}\hline\noalign{\smallskip}
        $\sum^{t_2}_{t_1}N_{SNN}(t)$ & 8.795$\times 10^6$&3.585$\times 10^6$\\
        $\sum^{t_2}_{t_1}N_{RNN}(t)$ &14.894$\times 10^6$&12.342$\times 10^6$\\
        $\sum^{t_2}_{t_1}N_{ANN}(t)$ & 1.065$\times 10^8$&1.065$\times 10^8$\\
        \noalign{\smallskip}\hline\hline
	\end{tabular}
\vspace{-12 pt}
\end{table}

\begin{subequations}
\begin{align}
    &    E_{SNN}(t)=E_{ACC}\cdot N_{ACC}^{SNN}(t)+E_{MAC}\cdot N_{MAC}^{SNN}(t)
        \label{energya}\\
    &    E_{RNN}(t)=E_{ACC}\cdot N_{ACC}^{RNN}(t)+E_{MAC}\cdot N_{MAC}^{RNN}(t)
        \label{energyb}\\
    &    E_{ANN}(t)=E_{ACC}\cdot N_{ACC}^{ANN}(t)+E_{MAC}\cdot N_{MAC}^{ANN}(t)
    \label{energyc}
\end{align}
\end{subequations}
{where, $E_{ACC}$  and $E_{MAC}$ are the energy cost of single
additions and multiplications respectively, which are respectively 0.1 pJ and 3.1 pJ\cite{Power_ana}. The total energy consumption of SNN and binary RNN from the integration of their power consumptions are also plotted in Fig. \ref{FIG_6}(e). The final energy consumption is presented in Table \ref{energycon}. With SNN staying \textit{idle}/consuming no energy when there is no spike, the hard sigmoid activation function for binary-activated RNN may still generate spikes, which makes it less energy efficient than SNN. As for ANN, it is always on with real floating values, consequently consuming much more energy than SNN and binary RNN.  
}\par

\subsubsection{Case II: Three-bus DC microgrid in ring topology}
As depicted in the three-bus system in Fig. \ref{FIG_5}(b), the same control objective in Case I is re-tested with more number of converters (with increased dimension of input data for training of SNN) and another topology. In Fig. \ref{FIG_7}(a) and (b), we compare the estimated currents and sampled currents, as well as the estimated voltages and sampled voltages. In that case, $\hat{i}_j^{(k)}$ represents the estimated $i_j$ by converter $k$. With increase in load current at $t_1$, the current sharing control objective is successfully met, since the estimated currents from SNN comply with the dynamic variations incurred by the remote measurements.\par
At the instant $t_3$, the voltage and current dynamics of the three buses are influenced by a line outage. Since the SNN has been trained using data that includes line outage scenarios, it is capable of recognizing the line outage and estimating the voltage and current of other converters. Consequently, $v_1$, $v_2$, and $v_3$ are regulated to re-distribute the power flow in the network, while $i_1$, $i_2$, and $i_3$ remain equal, demonstrating the capability of NSC-based control to operate under line outage conditions. At the time $t_4$, Converter III becomes completely disconnected, leading to the secondary control loop in Converter III ceasing its operation, as indicated by the dotted line. However, Converter I and Converter II continue to function and maintain current sharing.\par
With the outage of converter III in stage IV, the estimated currents $\hat{i}_j^{(k)}$ and voltages $\hat{v}_j^{(k)}$ in Fig. \ref{FIG_7}(a) and (b) closely match the corresponding measured values, demonstrating the effectiveness of NSC-based control in this case. 

Events in Fig. \ref{FIG_7}(c) arise from the dynamic process, and the corresponding spikes in Fig. \ref{FIG_7}(d) are generated in tandem with these events. The calculation of power consumption by SNN, RNN, and ANN (Fig. \ref{FIG_7}(e)) follows the same procedure as in Case I. The total energy consumption of SNN and binary RNN is detailed in Table \ref{energycon}. \par

\subsubsection{Case III: Two-bus DC microgrid with an intermediate solid-state transformer (SST)}
In this scenario in Fig. 4 (c), all the values of Converter II are referred to the primary side of the SST as a reference for comparison with a control objective on power sharing between converters. In Fig. \ref{FIG_9}(a) and (b), we compare the estimated currents $\hat{i}_1$, $\hat{i}_2$ with the sampled currents $i_1$ and $i_2$, as well as the estimated voltages $\hat{v}_1$, $\hat{v}_2$ with the sampled voltages $v_1$ and $v_2$ during dynamic disturbances.\par
It can be seen in Fig. \ref{FIG_9} that in both Stage I and Stage II, the estimated values $\hat{i}_2$ and $\hat{v}_2$ by Converter I closely match the sampled values $i_2$ and $v_2$ during dynamic processes, and vice versa. This results in equal power reference $P_1$ and $P_2$, which lead to real output powers $P_1$ and $P_2$ following their respective references to reach an equal steady state in Fig. \ref{FIG_9}(c). Beyond the governing limitations of TPC, we establish that NSC-based control is not limited by galvanic isolation along the transmission lines even at heterogeneous voltage levels.

\subsubsection{Case IV: Three-bus DC microgrid in star topology}
Considering a secondary control objective of power-sharing for a three-bus microgrid in a star topology in Fig. \ref{FIG_5}(d), when the load changes from 6.4 kW to 14.4 kW at Converter I at instant $t_1$, the power-sharing control is successfully met in Fig. \ref{FIG_10} following the same principle as in Case III. At $t_3$, a line outage occurs, resulting in voltage regulation to redistribute the power flow. This leads to equal power references, as shown in Fig. \ref{FIG_10}(c). At $t_4$, Converter III is disconnected, causing the secondary control loop in Converter I to isolate, as indicated by the dotted line. However, Converters I and II still resort to power sharing.\par
Despite the isolation of Converter III in Stage IV, the estimated currents $\hat{i}_j$ and voltages $\hat{v}_j$ in Fig. \ref{FIG_10}(a) and (b) closely match the measured values, respectively. Consequently, $P_1$, $P_2$, and $P_3$ are equal, resulting in accurate estimation of remote measurements and signals using the proposed NSC-based coordination.
\begin{figure}
    \centering
    \includegraphics[width=0.85\linewidth]{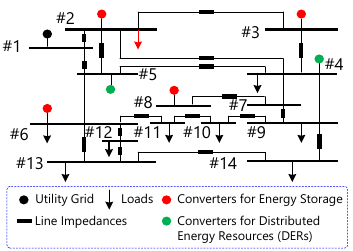}
    \caption{{Case V: Modified IEEE 14-bus DC system with energy storages and distributed energy resources (DERs).}}
    \label{IEEE14}
\end{figure}
 \begin{figure*}[thb]
        \centering
        \includegraphics[width=0.85\linewidth]{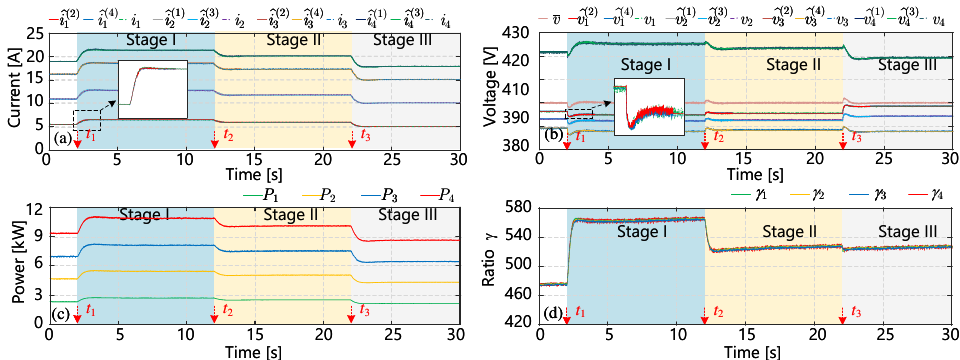}
        \caption{Simulation results of the ES converters: (a) comparison of estimated currents and respective measured values, (b) comparison of estimated voltages and respective measured values, (c) output powers, and (d) $\gamma$ ($P_i$/SOC ratio) of the ES converters.}
        \label{IEEE_result}
    \end{figure*}
\begin{figure}[h]
\centering
    \includegraphics[width=0.85\linewidth]{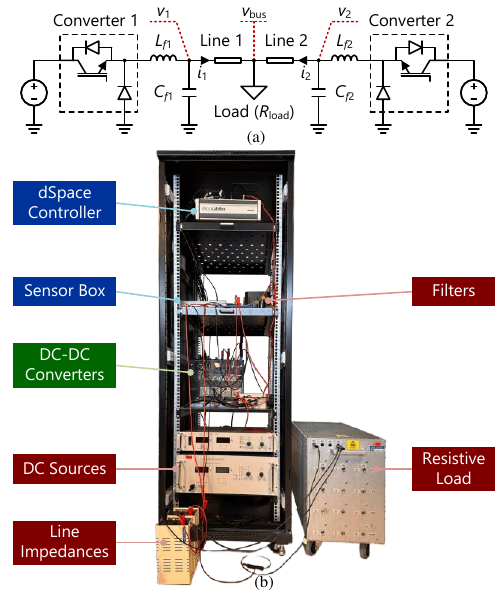}
    \caption{Experimental setup: (a) Single-line diagram of the two DC/DC converters tied to each other via a common resistive load, (b) picture of the prototype.}
    \label{Setup}
\end{figure}

\begin{figure*}[htb]
      \centering
        \includegraphics[width=0.71\linewidth]{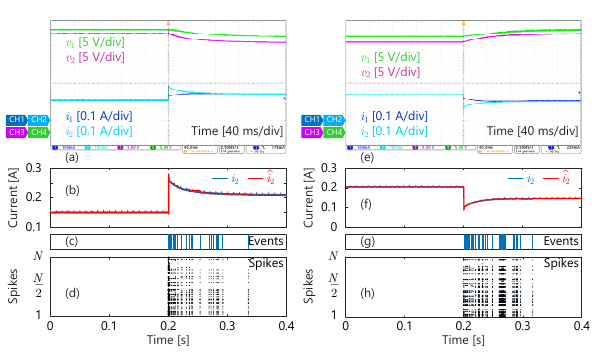}
        \caption{Experimental results: (a) voltage and current when load increases, (b) current of Converter II and its estimated value by Converter I when load increases, (c) events when load increases, (d) the spikes of hidden layer in SNN, (e) voltage and current when load decreases, (f) current of Converter II and its estimated value by Converter I when load decreases, (g) events when load decreases, (h) the spikes of hidden layer in SNN.}
        \label{loadchange}
\end{figure*}
\begin{figure*}[thb]
      \centering
        \includegraphics[width=0.71\linewidth]{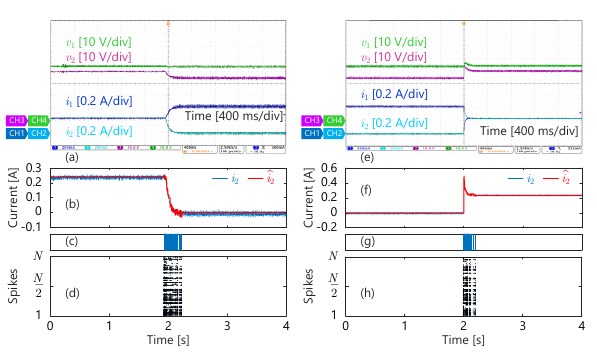}
        \caption{Experimental results: (a) voltage and current when Converter II quits, (b) current of Converter II and its estimated value by Converter I when Converter II quits, (c) events when Converter II quits, (d) the spikes of hidden layer in SNN, (e) voltage and current when Converter II resumes operation, (f) current of Converter II and its estimated value by Converter I when Converter II resumes operation, (g) events when Converter II resumes operation, (h) the spikes of hidden layer in SNN.}
        \label{lineout}
\end{figure*}
\subsubsection{Case V: A modified IEEE 14-bus system}
{To better validate the scalability of the proposed method in a large system, tests are also conducted in a modified IEEE 14-bus system in Fig. \ref{IEEE14}, of which the network structure is identical to the standard IEEE 14-bus system but with scaled down DC interconnection network and sources.} \par
{The DERs are working at their maximum power point and only controlled by respective local controllers. The energy storage (ES) converters connected to batteries need to collaborate with each other to support the dynamic change of generated power from the DERs. The coordinated control objectives of secondary control are average voltage regulation and proportionate power sharing. The average voltage is regulated at a rated voltage of 400 V. The power between the ES based converters is shared in a ratio of their respective state-of-charge (SOC) values, given by: }
\begin{equation}
    \gamma=\frac{P_{ES1}}{\Delta SOC_1}=\frac{P_{ES2}}{\Delta SOC_2}=\frac{P_{ES3}}{\Delta SOC_3}=\frac{P_{ES4}}{\Delta SOC_4}
             \label{22}
\end{equation}
such that    \begin{equation}
        \Delta SOC_{i}=\left\{
        \begin{array}{ll}
        SOC_{max}-SOC_i &\textit{when}\:\textbf{charging}\\
        SOC_i-SOC_{min} &\textit{when}\:\textbf{discharging}
         \end{array} \right. 
    \end{equation}
where, {$SOC_{max}$ and $SOC_{min}$ are the maximum and minimum SOC limits of the batteries. The real-time SOC of each ES can be estimated by the following equation, }
\begin{equation}
    SOC(t)=SOC_\mathrm{initial}-\frac{1}{C}\int i_\mathrm{bat}(t)\mathrm{d}t
\end{equation}
{The current $i_{bat}$ can be estimated by SNN, so the SOC of each ES can be known by the adjacent converters}.\par
{As we compare the estimated currents and voltages with the sampled values during dynamic disturbances, it can be seen in Fig. \ref{IEEE_result}(a) and (b) that in Stage I, Stage II, and Stage III, the estimated values closely match the sampled values, which results in the power-sharing results shown in Fig. \ref{IEEE_result}(c) and (d). They are aiming to be shared by the ratio of $\Delta SOC$. As shown in Fig. \ref{IEEE_result}(d),  the ratio $\gamma$ ($P_i$/$\Delta SOC$) is always equal, which indicates that the power $P_1$ - $P_4$ can always be shared according to the ratio of $\Delta SOC$ in (\ref{22})}. 
    
\subsection{Experiment results}
{A two-bus system illustrated in Fig. \ref{Setup}(b) has been employed for experimental validations, with the single-line diagram and other parametric details presented in Fig. \ref{Setup}(a). It comprises two DC/DC buck converters managing equal load current sharing. Dynamic events, such as load change and line outage are considered to check the efficacy of NSC under real-time conditions. The experimental parameters can be found in Table \ref{ex_para}. The offline design of SNN and its corresponding parameters can be found in Appendix.}
\begin{table}[tbh]
    \centering
    \caption{{Experimental Parameters.}}
    \label{ex_para}
    \begin{tabular}{c c c} \hline \hline 
        \textbf{Parameter}&  \textbf{Symbol}&\textbf{Specification}\\ \hline 
        Rated voltage& $V_n$&40 V\\ 
        Rated power&  $P_1$=$P_2$&50 W\\
  Filter inductance& $L_{f1}$=$L_{f2}$&1.5 mH\\
  Filter capacitance& $C_{f1}$=$C_{f2}$& 700 \textmu F\\ 
          Line resistance& $ R_{1}$, $ R_{2}$&1.5 \textOmega, 3.6 \textOmega\\ 
          Load resistance& $ R_\mathrm{load}$&115 \textOmega\\ 
        \hline\hline
    \end{tabular}

\end{table}
\subsubsection{Load change}

In Fig. \ref{loadchange}(a), before the load change, currents $i_1$ and $i_2$ are equal. When the load transitions from 115 \textOmega\ to 75 \textOmega at t=0.2 s, a dynamic process is formalized leading to generation of output events, such that $i_1$ and $i_2$ maintain equal sharing. The voltage variables $v_1$ and $v_2$ experience a sag due to increased load, reaching two new steady states. In Fig. \ref{loadchange}(b), $i_2$ is compared with its estimated value $\hat{i}_2$ by Converter I, ensuring accurate current sharing. During the dynamic process, events are generated, as depicted in Fig. \ref{loadchange}(c), leading to corresponding spikes in the hidden layer of SNN, presented in Fig. \ref{loadchange}(d).\par
Fig. \ref{loadchange}(e) illustrates the load recovery from 75 \textOmega\ to 115 \textOmega, with both $i_1$ and $i_2$ decreasing and sharing the load equally after a dynamic process. Voltage variables $v_1$ and $v_2$ increase and return to their initial steady states. Control system variables for Converter I in this scenario are shown in Fig. \ref{loadchange}(f) and (g). As seen in Fig. \ref{loadchange}(f), the estimated value $\hat{i}_2$ accurately matches $i_2$, ensuring precise current sharing. During the dynamic process, events are generated in (g), and corresponding spikes in the hidden layer of SNN are generated in Fig. \ref{loadchange}(h) synchronized with the events.\par
\subsubsection{Line outage}
In Fig. \ref{lineout}(a), before the line outage, $i_1$ and $i_2$ share an equal load. When Converter II is disconnected due to a line outage, both $v_1$ and $v_{\text{bus}}$ sag due to power droop control. In Fig. \ref{lineout}(b), $i_2$ is compared with its estimated value $\hat{i}_2$ by Converter I, enabling Converter I to be aware of the states of Converter II. As Converter I operates in islanded mode, NSC based secondary control becomes inactive. During the dynamic process, events are generated, as depicted in (c), leading to corresponding spikes presented in Fig. \ref{lineout}(d).\par
Fig. \ref{lineout}(e) illustrates the scenario when Converter II is reconnected to the system at t = 2 s, with $i_1$ and $i_2$ returning to an equal state. This occurs because SNN is activated by local dynamics, allowing it to estimate $i_2$ again, as shown in Fig. \ref{lineout}(f). During the dynamic process, the SNN is activated by the events in Fig. \ref{lineout}(g), with corresponding spikes shown in Fig. \ref{lineout}(h).\par

\section{Conclusions and Future scope of work}
This paper explores a low-power neuromorphic inference based application for the first time in the realm of microgrids and power systems. We firstly leverage the deterministic features of information-theoretic learning using the highly energy-efficient spiking neural networks (SNNs) at each bus to estimate the remote measurements only using the unmodulated power flows as the information carrier to enable co-transfer of power and information. Secondly, we explore its application for the hierarchical control of microgrids as its application {by leveraging the \textit{publish-subscribe} architecture as a novel communication protocol for power electronic systems}. Thirdly, since SNNs are trained using spikes based binary data, we translate an event-driven sampling process to reflect on the most significant communicative signatures from the remote ends that are further converted to spikes for online training and inferences from SNNs, that is updated with every system transients. By doing so, we not only eliminate the communication infrastructure and their associated reliability and security concerns, but also clearly demonstrate the computational advantages, energy efficiency and feasibility behind using SNN over other data and computational-intensive neural networks. {The proposed framework also eliminates the cyber layer vulnerabilities to restrict any exogenous path arrivals for the cyber attackers. Furthermore, it doesn't suffer from the inefficiency and scalability issues that other co-transfer technologies such as \textit{Talkative Power Communication} will incur.} \par
In the future, we plan to expand this philosophy to facilitate smarter integration or start-up of distributed energy sources and regional microgrids. The proposed scheme still needs to be validated on fully parallel memristor crossbar array circuits based processors with a specific focus on accuracy and versatility for different noise levels.

\appendix
\subsection{SNN Parameters -- Simulation Studies}
Number of hidden layers = 2, Number of neurons in encoding and hidden layer = 256, Number of neurons in decoding layer = 4, $\sigma^{V}_{th}$ = 0.01, $\sigma^{I}_{th}$ = 0.002, $\sigma^{I}_{th}$ = 0.0039.\par
Dataset dimensions for the inputs and outputs in Case I and III : $\mathbb{D}_{in,800\times 4}\:=\:\{\bm{v}_{i,800\times 1},\:\bm{i}_{i,800\times 1},\:\bm{\dot{v}}_{i,800\times 1},\:\bm{\dot{i}}_{i,800\times 1}\}$, {\mbox{$\mathbb{D}_{out,800\times 2}\:=\:\{\bm{\hat{v}}^{(i)}_{j,800\times 1},\:\bm{\hat{i}}^{(i)}_{j,800\times 1},\}$}}, 
$\forall \ i,j \in \{1,2\}, i\neq j$.

Dataset dimensions for the inputs and outputs in Case II and IV : $\mathbb{D}_{in,800\times 4}\:=\:\{\bm{v}_{i,800\times 1},\:\bm{i}_{i,800\times 1},\:\bm{\dot{v}}_{i,800\times 1},\:\bm{\dot{i}}_{i,800\times 1}\}$, {\mbox{$\mathbb{D}_{out,800\times 4}\:=\:\{\bm{\hat{v}}^{(i)}_{j,800\times 1},\:\bm{\hat{i}}^{(i)}_{j,800\times 1},\:\bm{\hat{v}}^{(i)}_{k,800\times 1},\:\bm{\hat{i}}^{(i)}_{k,800\times 1}\}$}}, 
$\forall \ i, j, k \in \{1, 2, 3\}, i\neq j\neq k$.
\subsection{SNN Parameters -- Experimental Studies}
Number of hidden layers = 2, Number of neurons in encoding and hidden layer = 64, Number of neurons in decoding layer = 4, $\sigma^{V}_{th}$ = 0.41, $\sigma^{I}_{th}$ = 0.0063, $\sigma^{I}_{th}$ = 0.024.\par
Dataset dimensions for the inputs and outputs: $\mathbb{D}_{in,4000\times 4}\:=\:\{\bm{v}_{i,4000\times 1},\:\bm{i}_{i,4000\times 1},\:\bm{\dot{v}}_{i,4000\times 1},\:\bm{\dot{i}}_{i,4000\times 1}\}$, {\mbox{$\mathbb{D}_{out,4000\times 2}\:=\:\{\bm{\hat{v}}^{(i)}_{j,4000\times 1},\:\bm{\hat{i}}^{(i)}_{j,800\times 1},\}$}}, 
$\forall \ i,j \in \{1,2\}, i\neq j$.


%\section*{Acknowledgment}


\ifCLASSOPTIONcaptionsoff
  \newpage
\fi



%\bibliographystyle{IEEEtran}

\bibliography{Reference.bib}
\bibliographystyle{IEEEtran}

%\end{IEEEbiography}


%\newpage



%\vfill


\end{document}